\documentclass[iop,apj,tighten]{emulateapj}
\usepackage{apjfonts}
\usepackage[breaklinks,colorlinks,citecolor=blue,linkcolor=magenta]{hyperref}

\shorttitle{The Eating Habits of Milky Way Mass Halos}
\shortauthors{Deason, Mao \& Wechsler}

\begin{document}

\title{The Eating Habits of Milky Way Mass Halos: destroyed dwarf satellites and the metallicity distribution of accreted stars}

\author{Alis~J.~Deason, Yao-Yuan~Mao, Risa~H.~Wechsler}
\affil{Kavli Institute for Particle Astrophysics and Cosmology \& Physics Department, Stanford University, Stanford, CA 94305, USA\\
SLAC National Accelerator Laboratory, Menlo Park, CA 94025, USA\\
{\tt adeason@stanford.edu}}

\begin{abstract}
We study the mass spectrum of destroyed dwarfs that contribute to the accreted stellar mass of Milky Way (MW) mass ($M_{\rm vir} \sim 10^{12.1} M_\odot$) halos using a suite of 45 zoom-in, dissipationless simulations. Empirical models are employed to relate (peak) subhalo mass to dwarf stellar mass, and we use constraints from $z=0$ observations and hydrodynamical simulations to estimate the metallicity distribution of the accreted stellar material. The dominant contributors to the accreted stellar mass are relatively massive dwarfs with $M_{\rm star} \sim 10^8-10^{10}M_\odot$. Halos with more quiescent accretion histories tend to have lower mass progenitors ($10^{8}-10^9 M_\odot$), and lower overall accreted stellar masses. Ultra-faint mass ($M_{\rm star} < 10^5M_\odot$) dwarfs contribute a negligible amount ($\ll 1\%$) to the accreted stellar mass and, despite having low average metallicities, supply a small fraction ($\sim2-5\%$) of the very metal-poor stars with [Fe/H] $< -2$. Dwarfs with masses $10^5 < M_{\rm star}/M_\odot < 10^8$ provide a substantial amount of the very metal-poor stellar material ($\sim40-80\%$), and even relatively metal-rich dwarfs with $M_{\rm star} > 10^8M_\odot$ can contribute a considerable fraction ($\sim20-60\%$) of metal-poor stars if their metallicity distributions have significant metal-poor tails. Finally, we find that the generic assumption of a quiescent assembly history for the MW halo seems to be in tension with the mass spectrum of its surviving dwarfs. We suggest that the MW could be a ``transient fossil''; a quiescent halo with a recent accretion event(s) that disguises the preceding formation history of the halo.

\end{abstract}

\keywords{Galaxy: formation --- Galaxy: halo ---  Galaxy: stellar content --- galaxies: dwarf --- galaxies: interactions --- Local Group}

\section{Introduction}
Dark matter halos grow hierarchically over time from the aggregation of several lower mass ``subhalos''. The rate of growth, and the mass spectrum of lower mass progenitors, strongly depend on the mass of the host halo, as well as its surrounding environment. However, even at fixed halo mass, the halo-to-halo scatter is large, reflecting the breadth of different assembly histories shaping each dark matter halo. Relating the build up of cold dark matter to the growth of stellar material in galaxy halos is non-trivial; the relation between stellar and dark matter mass is highly non-linear (e.g., \citealt{conroywechsler}; \citealt{behroozi13c}; \citealt{moster13}). For example, at low masses the $M_{\rm star}-M_{\rm halo}$ relation is very steep, and it is likely that below some mass threshold star formation is completely suppressed and subhalos are simply ``dark'' (e.g., \citealt{bullock00}; \citealt{benson02}; \citealt{kravtsov04}).

The growth of stellar mass in Milky Way (MW) mass halos is generally dominated by intrinsic star formation in the very center of the halo. However, stars can also be accreted from the digestion of lower mass subhalos that have their own stellar populations. While this accreted component is generally lower in mass than the stars born \textit{in-situ} \citep[e.g.][]{behroozi13c}, the stellar material splayed out throughout the halo is a remnant of the halo's assembly history, and provides a visual (as opposed to dark) record of the lower mass fragments that have contributed to the halo's growth over time.

Several studies have attempted to connect the predictions of the $\Lambda$CDM paradigm to the vast stellar halos that surround galaxies like the MW (e.g., \citealt{bullock05}; \citealt{purcell07}; \citealt{delucia08}; \citealt{cooper10}). A general consensus from these theoretical studies is that the majority of the stellar material accreted by MW mass halos comes from early, massive accretion events. However, studies specifically focusing on the build-up of ``MW-type'' halos are often biased to halos with quiescent accretion histories. There is significant evidence that our Galaxy has been largely undisturbed over the past several Gyr (e.g., \citealt{gilmore02}; \citealt{hammer07}), but this bias limits our ability to understand the true breadth in assembly histories at MW mass scales, and how this relates to the mass spectrum of accreted dwarfs. Furthermore, approximately $\sim 70\%$ of MW mass halos likely host disk galaxies (e.g., \citealt{weinmann06}; \citealt{choi07}). Thus, while restricting to quiescent accretion histories likely excludes most elliptical galaxies undergoing recent major mergers, a significant number of halos hosting disk galaxies are likely also excluded (cf.\ \citealt{stewart08}). This limitation is important if we want to place our own Galaxy's accretion history in context with other, similar mass disk galaxies.

In our own Galaxy, the chemical properties of halo stars have often been used to connect them to their progenitor galaxies. For example, the relation between [$\alpha$/Fe] and [Fe/H] can be linked to the host galaxy's mass (e.g., \citealt{tolstoy09}). Halo stars are typically more $\alpha$-enhanced at metallicities [Fe/H] $\gtrsim -1.5$ than the (classical) dwarf galaxy satellites in the MW (e.g., \citealt{tolstoy03}; \citealt{venn04}), which suggests that their progenitors are not represented in the surviving dwarf galaxy population. However, \cite{robertson05} (see also \citealt{font06}) showed that this mismatch in chemical properties can be reconciled if the progenitors of halo stars are biased towards early, massive accretion events, as predicted from $\Lambda$CDM simulations. More massive progenitors have also been favored from recent observational studies. \cite{deason15} showed that the (relatively high) ratio of blue straggler to blue horizontal branch stars in the stellar halo favors more massive dwarfs as the ``building blocks'' of the stellar halo, and \cite{fiorentino15} found that the period and luminosity amplitudes of RR Lyrae stars in the halo are more consistent with massive dwarfs than lower-mass dwarfs.

There is a fair amount of agreement, at least qualitatively, between observations and theory that relatively massive dwarf galaxies are the dominant contributors to the overall accreted stellar material in MW mass galaxies. However, it is less clear what mass progenitors supply the majority of the very metal-poor ([Fe/H] $\lesssim -2$) material in the halo. Early studies of the metallicity distributions of the classical dwarfs found a lack of very metal-poor stars in these galaxies (e.g., \citealt{helmi06}), however a re-calibration of the Calcium \textsc{ii} triplet lines at low metallicities (e.g., \citealt{starkenburg10}) found that these dwarfs are not as devoid of low-metallicity stars as previously thought, and their metal-poor tails are similar to the MW halo stars. In fact, the abundance ratios of metal-poor stars in classical dwarfs are indistinguishable from the halo population (e.g., \citealt{tolstoy09}).  Moreover, the ``ultra-faint'' dwarf galaxy population ($M_{\rm star} \lesssim 10^5M_\odot$) has similar chemical properties as metal-poor halo stars (e.g., \citealt{frebel10}; \citealt{norris10}; \citealt{simon10}; \citealt{lai11}) and very low average metallicities (e.g., \citealt{kirby13}). It has been suggested that these low mass ultra-faints could be the dominant source of the very metal-poor stellar material in the MW (e.g., \citealt{frebel10}). However, cosmological models quantifying the contribution of ultra-faint mass dwarfs to the accreted stellar component of galaxies are scarce. High resolution simulations are needed to resolve down to these mass scales, and the relation between subhalo mass and stellar mass at these low-mass scales is still rather uncertain (cf.\ recent theoretical determinations, e.g., \citealt{gk14}; \citealt{hopkins14}). Furthermore, while the abundance of low mass subhalos in dark matter only simulations is very high, it is likely that only a small fraction of these subhalos host luminous galaxies (e.g., \citealt{sawala14}).

It is clear that there is some bias in comparing the surviving dwarf galaxy population with the dwarfs that were destroyed several Gyr ago. The survivors are generally lower mass and have likely experienced more prolonged star formation than their destroyed counterparts. However, the mass spectrum of surviving satellite galaxies at the group/cluster mass-scale has often been related to the assembly histories of their halos (e.g., \citealt{conroy07}; \citealt{dariush10}; \citealt{deason13b}). For example, the ``magnitude-gap'' statistic, the difference in absolute magnitude between the most massive satellite galaxy and the central galaxy, is often used to distinguish old, quiescent halos (i.e., fossil groups) from groups undergoing recent major mergers. However, on galaxy scales the relation, if any, between the halo assembly histories and the mass spectrum of surviving dwarfs is relatively unexplored. Furthermore, if both the surviving and destroyed dwarfs are signposts of halo accretion histories, we should be more invested, both observationally and theoretically, in finding a link between the dwarfs that survived and those that perished.

In this contribution, we employ a suite of 45 zoom-in simulations to investigate the mass spectrum of destroyed dwarfs that contribute to the accreted stellar mass of MW mass halos. Our simulation suite spans a narrow mass range ($M_{\rm vir} \sim 10^{12.1 \pm 0.03} M_\odot$), but has a wide range of accretion histories. This allows us to focus solely on the relation between halo assembly history and the growth of stellar mass from accreted lower mass fragments. Furthermore, the high-resolution of these zoom-in simulations allows us to study subhalos down to the ultra-faint mass dwarf scale in a fully cosmological context. The paper is arranged as follows. Section \ref{sec:sims} describes our simulation suite and outlines how we assign stellar mass to dark matter subhalos. In Section \ref{sec:mspec} we investigate the stellar mass spectrum of destroyed dwarfs in MW mass galaxies, and relate the dominant contributors of the accreted stellar mass to the host halo assembly histories. We use constraints from $z=0$ observations and hydrodynamical simulations in Section \ref{sec:met} to predict the metallicity distribution of the accreted stellar material, and we estimate the contribution of different mass dwarfs to the metal-poor stellar component. In Section \ref{sec:sats} we relate the most massive surviving dwarf satellite galaxies to the mass spectrum of destroyed dwarfs. Finally, we summarize our main conclusions in Section \ref{sec:conc}.

\section{Simulations}
\label{sec:sims}

\begin{figure}
\begin{center}
\includegraphics[width=8.5cm, height=9.9cm]{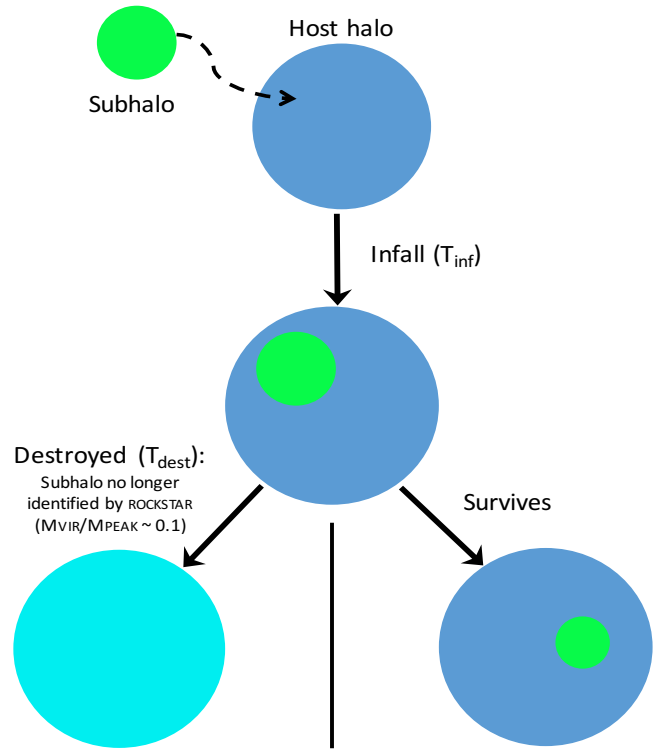}
\caption{\label{fig:cartoon} \small A schematic illustration of our definition of ``destroyed'' and ``surviving'' subhalos. Destroyed subhalos typically lose $\sim$90\% of their peak mass before they are no longer identified by the \textsc{Rockstar} halo finder. Surviving subhalos are still identified by \textsc{Rockstar} at $z=0$.}
\end{center}
\end{figure}

We use a suite of 45 zoom-in simulations of MW mass halos (\citealt{mao15}). The halos are selected from a low-resolution dark matter only cosmological simulation (\texttt{c125-1024} box) with cosmological parameters $\Omega_m=0.286$, $\Omega_\Lambda=0.714$, $h=0.7$, $\sigma_8=0.82$, and $n_s=0.96$. The low-resolution box has $1024^3$ particles with a side length of 125 Mpc$\,h^{-1}$, and was run using \textsc{L-Gadget} (\citealt{springel01,springel05}).

The selected halos fall in the narrow mass-range $M_{\rm vir}=10^{12.1\pm 0.03}M_\odot$ in the \texttt{c125-1024} box. The initial conditions of the zoom-in
simulations are generated using \textsc{Music} (\citealt{hahn11}), and the Lagrangian volume enclosing the highest-resolution particles is set by the rectangular volume that the particles within $10 R_{\rm vir}$ of the $z=0$ halo occupied at $z=99$. The mass resolution in the zoom regions is $3.0 \times 10^5M_\odot h^{-1}$, and the softening length is 170 pc$\,h^{-1}$ comoving. Note that the zoom-in simulations are not randomly chosen from the mass-selected sample in the \texttt{c125-1024} box. The zoom-ins are slightly biased towards early forming halos, but span a wide range of accretion histories.

Dark matter subhalos are identified using the six-dimensional halo finder \textsc{Rockstar} (\citealt{behroozi13a}), and merger trees were constructed by \textsc{Consistent Trees} (\citealt{behroozi13b}) with 235 snapshots between $z=0$ and 19, equally spaced in logarithmic scale factor. Halos are assigned a virial mass, $M_{\rm vir}$, and radius, $R_{\rm vir}$, using the evolution of the virial relation from \cite{bryan98}. For our cosmology this corresponds to an overdensity of $\Delta_{\rm crit}=99.2$ at $z=0$. Host halos are defined as isolated halos than can host lower mass subhalos within their virial radii, and subhalos are defined as halos that are within $R_{\rm vir}$ of a more massive host halo. We compute the peak mass, $M_{\rm peak}$ of each subhalo as the maximum mass that a subhalo ever reached along the main branch of its progenitor.

\subsection{Identifying Destroyed Subhalos}
We trace back the progenitors of host halos at each simulation time step, and identify all progenitors that are \emph{not} the most massive progenitor as the ``destroyed'' (sub)halos. In other words, when a (sub)halo is no longer tracked by the halo finder, it is considered ``destroyed''. All other subhalos are 
``surviving'' subhalos, i.e., those subhalos which are tracked by the halo finder down to $z=0$. A cartoon illustrating our definition of ``destroyed'' and ``surviving'' subhalos is shown in Figure \ref{fig:cartoon}. With the above definition, it is clear that the time when a subhalo is destroyed depends on how the halo finder identifies subhalos and how the tree builder links progenitors. For example, adjusting the parameter ``unbound threshold'' in \textsc{Rockstar} would affect how long a merged, stripped subhalo is tracked. As a sanity check, we apply the iterative unbinding procedure in \textsc{Rockstar} developed by \cite{griffen15} to a high resolution (particle mass $= 4 \times 10^4M_\odot h^{-1}$) version of one of the host halos (Halo 937). We find that the inclusion of iterative unbinding increases the median mass-loss of subhalos with $M_{\rm peak} > 10^{8} M_\odot$ before destruction to 97\% (cf. 90\% for the fiducial runs). However, we find that the inclusion of this algorithm does not significantly affect our main results.
Different halo finders and tree builders could also produce different results; we refer the reader to \citet{avila2014} for a detailed comparison.
Nevertheless, in our analysis, the conservative resolution criterion we have applied helps to minimize the impact of these uncertainties. 

Subhalos can lose a significant amount of mass ($\sim 90\%$) before they are ``destroyed", but this definition of destruction is a good proxy for when the \textit{stellar mass} associated with a subhalo is liberated into the host halo (see e.g., \citealt{penarrubia08}; \citealt{wetzel10}).
Note, however, that the true definition of when the stellar material from subhalos is liberated into the main halo is highly uncertain, and likely dependent on the orbital properties and mass of the subhalos as well as the subhalo finder used in the analysis. Moreover, even observationally, it is unclear when a dwarf undergoing tidal stripping should no longer be identified as a distinct object (cf.\ the Sagittarius dwarf in the MW). In our analysis, we use the simple definition described above for destroyed subhalos and focus on the relative differences between host halos, however, it is worth bearing in mind that the derived \textit{time} of subhalo destruction does depend on our adopted definition.

In our analysis, we only consider subhalos that are progenitors of the \textit{host halo}. Thus, we do not take into account ``sub-subhalos'' that can be destroyed within the virial radius of the progenitor subhalos before they themselves are destroyed. The population of sub-subhalos can be significant, especially at the low-mass end (e.g \citealt{wetzel15a}). To investigate the potential effect of this population, we track each host halo progenitor back to its (first) infall onto the host halo, and consider its own subhalo population at infall. Those sub-subhalos that get destroyed within the virial radius of the progenitor subhalos after infall onto the host halo can be counted as additional progenitors of the host halo. However, we find that the inclusion of this population makes little difference to our results, so we do not include these destroyed sub-subhalos in the remainder of the analysis. 

Note that in this work, we do not consider the subhalos of progenitors that are destroyed \textit{before} these progenitors fall into the host halos. For example, a massive dwarf that is eventually destroyed in a MW-mass halo has its own accretion history while it is an isolated halo (i.e., before infall), and several smaller mass dwarfs may have contributed to the mass of this massive dwarf. In this study, we only consider the mass-spectrum of subhalos destroyed within the virial radius of the main (MW-mass) host halo, which, by definition, excludes any subhalos or sub-subhalos destroyed before infall onto the host.

Throughout our analysis, we only consider subhalos with $M_{\rm peak} > 10^8M_\odot$ ($V_{\rm max} \gtrsim 9$ km s$^{-1}$). \cite{mao15} estimate that this a conservative lower limit for convergence of the zoom-in MW simulations. For one of our host halos (Halo 937), we have a higher resolution run (particle mass $= 4 \times 10^4M_\odot h^{-1}$) that we can use to test for numerical convergence. By directly comparing this higher resolution simulation with its lower resolution counterpart (particle mass $= 3 \times 10^5M_\odot h^{-1}$), we confirm that our results are robust to numerical resolution effects.

\subsection{Assigning Stellar Mass to Subhalos}
\label{sec:am}

We assign stellar mass to subhalos using the $M_{\rm star}-M_{\rm peak}$ relation derived by \cite{gk14}; these authors showed that this relation agrees well with number counts of $z=0$ local group dwarfs. We apply 0.2 dex scatter in log $M_{\rm star}$ at fixed log $M_{\rm peak}$ for $M_{\rm peak} > 10^{11}M_\odot$ and 0.3 dex scatter for lower mass subhalos with $M_{\rm peak} < 10^{11}M_\odot$. Our lower mass threshold of $M_{\rm peak} > 10^8M_\odot$ for subhalos corresponds to a stellar mass limit of $M_{\rm star} \gtrsim 10^{2.6}M_\odot$.

 We assume no redshift evolution in the $M_{\rm star}-M_{\rm peak}$ relation. This assumption is motivated by the lack of evidence for a strong redshift evolution on dwarf mass-scales from either theoretical (\citealt{hopkins14}; \citealt{graus15}), empirical \cite[e.g.][]{behroozi13c}, or observational (\citealt{wake11}; \citealt{leauthaud12}; \citealt{hudson13}) studies. However, we do check that employing the redshift-dependent $M_{\rm star}-M_{\rm halo}$ relations derived by \cite{behroozi13c} and \cite{moster13} makes little difference to our main results. Note that when we adopt these redshift-dependent relations we use the mass and redshift \textit{at infall} onto the host halo when assigning stellar mass to subhalos.

Our prescription assumes that \textit{all} subhalos down to $M_{\rm peak} \sim 10^8M_\odot$ (the resolution limit of the simulations) host a central (dwarf) galaxy. However, several studies (e.g., \citealt{okamoto09}; \citealt{nickerson11}; \citealt{shen14}; \citealt{sawala14}) have shown that reionization can prevent star formation in halos below $\sim 10^{9.5}M_\odot$ ($M_{\rm star} \sim 10^5M_\odot$). Thus, at the ultra-faint dwarf mass scale, not all subhalos will form stars. Our implementation will therefore \textit{overestimate} the number of ultra-faint dwarfs in the simulations, and thus their contribution to the accreted stellar mass (see Section \ref{sec:fracs}) is likely an upper limit.

Note that we use the term ``dwarf'' to describe the stellar component of \textit{all} subhalos with $M_{\rm peak} > 10^8M_\odot$. This includes a small number galaxies with stellar masses ($M_{\rm star} \gtrsim 10^{9}M_\odot$) that would not normally be considered as dwarf galaxies. For our purposes, we use this loose definition of ``dwarf'' to describe the galaxies less massive than the main host throughout the paper, but caution the reader that this definition does include a handful of more massive galaxies (particularly in recent major mergers).

In the following sections, we consider the stellar mass that is accreted by MW mass halos from dwarf progenitors\footnote{Note that this does not include the stellar mass residing in surviving $z=0$ dwarf satellites.}. We note that this does not include any of the stellar mass born \textit{in situ} in the central galaxy, which generally comprises the majority of the stellar mass budget on these mass scales. Furthermore, the accreted stellar mass need not reside solely in the galaxy stellar halos, as a significant fraction can end up in the disk/bulge (\citealt{read08}; \citealt{cooper10}; \citealt{pillepich15}). Where possible, we make approximate comparisons with observations of stellar halos, but note that \textit{direct} comparisons should be taken with a healthy grain of salt. 

Throughout this work we consider ``mass-weighted'' quantities, which will naturally bias us towards the inner regions of observed halos ($\lesssim 20$ kpc). However, while it is beyond the scope of this work to probe radial trends in galaxy halos (see e.g., recent work by \citealt{amorisco15} and \citealt{gomez15}), we do focus on the region where most of the accreted stellar mass resides.  

\section{Mass Spectrum of Destroyed Dwarfs}
\label{sec:mspec}

\begin{figure}
\begin{center}
\includegraphics[width=8.5cm, height=7.08cm]{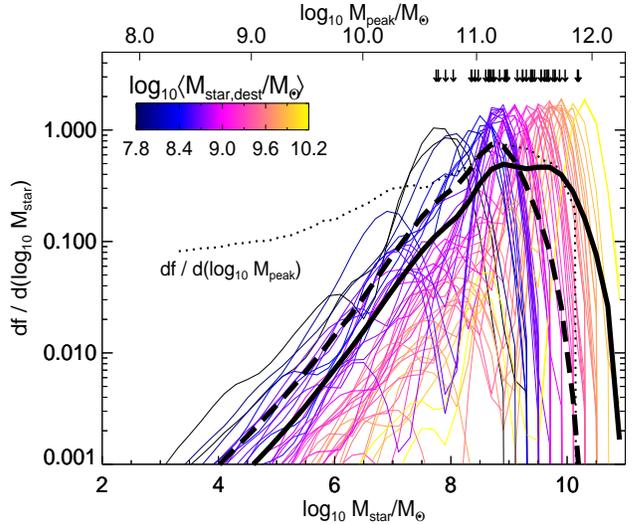}
\caption{\label{fig:mspec_mstar} \small The differential contribution to the accreted stellar mass from destroyed dwarfs with stellar mass $M_{\rm star}$ as a function of $M_{\rm star}$. The thin colored lines show the differential mass fractions for each of the 45 host halos. The colors indicate the average mass of the dwarf contributors, weighted by stellar mass (darker/blue = low masses, lighter/yellow = high masses). Median values for each host distribution are indicated by the arrows, these range from $M_{\rm star} \sim 10^{8} - 10^{10}M_\odot$. The thick solid black line shows the average profile over all 45 host halos. The thick dashed black line shows the average profile for the sample of 19 ``quiescent'' host halos, which have not undergone a major merger since $z=2$; here, the differential mass contribution is biased towards lower dwarf masses. We also show the differential contribution of peak subhalo masses ($M_{\rm peak}$) with the dotted line. This profile differs from the stellar mass owing to the steep decline in the stellar mass--halo mass relation at low halo masses ($M_{\rm peak} \lesssim 10^{11}M_\odot$).}
\end{center}
\end{figure}

In this Section, we investigate the ``mass-spectrum'' of destroyed dwarfs of our 45 MW-mass halos. We show the differential contribution to the accreted stellar mass from destroyed dwarfs with stellar mass $M_{\rm star}$ as a function of $M_{\rm star}$ in Figure \ref{fig:mspec_mstar}. 

The mass weighted average stellar mass for destroyed dwarfs gives the typical dwarf mass of contributors to the total accreted stellar mass.
\begin{equation}
\label{eq:mav}
\langle M_{\rm star, dest} \rangle = \frac{\sum_i M^i_{\rm star, dest} \times M^i_{\rm star, dest}}{\sum_i M^i_{\rm star, dest}}
\end{equation}
Here, the sum is performed over all destroyed dwarfs with $M_{\rm peak} > 10^8M_\odot$.
Similarly, we can define the mass weighted average time when these dwarfs are destroyed:
\begin{equation}
\langle T_{\rm dest} \rangle = \frac{\sum_i T^i_{\rm dest} \times M^i_{\rm star, dest}}{\sum_i M^i_{\rm star, dest}}
\end{equation}

The thin colored lines in Figure \ref{fig:mspec_mstar} show the differential contributions to the accreted stellar mass for each host halo. The lines are colored according to $\langle M_{\rm star, dest} \rangle$ (darker/blue = low mass, lighter/yellow = high mass), and $\langle M_{\rm star, dest} \rangle$ for each host halo is indicated by the black arrows. The typical $\langle M_{\rm star, dest} \rangle$ ranges from $10^8-10^{10} M_\odot$, but the halo-to-halo scatter is large.

The thick black line shows the average distribution. The dotted line shows the differential contribution of dark matter $M_{\rm peak}$. The steep $M_{\rm star}-M_{\rm peak}$ relation at low mass scales ($M_{\rm star} \propto M_{\rm peak}^{1.9}$) suppresses the contribution of subhalos with low $M_{\rm peak}$ to the accreted stellar mass.

We also define a subsample of host halos with a ``quiescent'' accretion history, defined as having no major mergers (dark matter mass ratio $> 0.3$) since $z=2$. Note that we only consider major mergers with dwarfs that are now destroyed; surviving satellites are not included. Only 40 \% (19) of the whole sample pass this cut. This quiescent criteria is generally used to define samples of ``Milky Way type'' halos in simulations (e.g., \citealt{bullock05}; \citealt{delucia08}; \citealt{cooper10}). However, while there is plenty of observational evidence suggesting that the MW has undergone a relatively quiescent accretion history (e.g., \citealt{gilmore02}; \citealt{hammer07}; \citealt{deason13a, deason14};  \citealt{ruchti15}), approximately $\sim 70\%$ of halos with $M_{\rm halo} \sim 10^{12}M_\odot$ are expected to host disk galaxies (e.g., \citealt{weinmann06}; \citealt{choi07}). Thus, the strict criteria for a quiescent accretion history likely excludes several halos that could host disk galaxies with similar mass to the MW. 

The distribution for the quiescent sample is shown with the thick dashed black line in Figure \ref{fig:mspec_mstar}; this sample is biased to lower $\langle M_{\rm star, dest} \rangle$ values, typically $\lesssim 10^9 M_\odot$. The typical mass dwarfs that are accreted by these quiescent halos are in good agreement with the findings of previous works attempting to model stellar halos of MW type galaxies (e.g., \citealt{bullock05}; \citealt{delucia08}; \citealt{cooper10}). 

\begin{figure}
\begin{center}
\includegraphics[width=8.5cm, height=4.25cm]{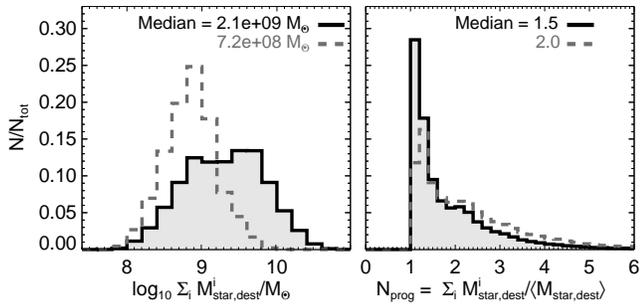}
\caption{\label{fig:nprog} \small \textit{Left panel:} The distribution of total stellar masses contributed by destroyed dwarfs. The median value for all 45 host halos is $2 \times 10^9 M_\odot$, but the halo-to-halo scatter is large (ranging from $10^8 - 10^{10.5}M_\odot$). The gray dashed line shows the distribution for the 19 quiescent host halos; these generally have lower accreted stellar masses. \textit{Right panel:} An estimate of the number of dwarf progenitors that contribute to the total accreted stellar mass, defined by $N_{\rm prog}=\sum_i M^i_{\rm star, dest}/\langle M_{\rm star, dest} \rangle$. Typically, $1-2$ destroyed dwarfs deposit the majority of accreted stellar mass onto the host halos.}
\end{center}
\end{figure}

The left-hand panel of Figure \ref{fig:nprog} shows the distribution of total stellar masses contributed by destroyed dwarfs for each host halo. The halo-to-halo scatter is large, with total stellar masses ranging from $10^8M_\odot$ to $10^{10.5}M_\odot$ (cf.\ \citealt{cooper13}). The quiescent halo sample generally has lower total accreted stellar masses. The right-hand panel of Figure \ref{fig:nprog} shows the typical number of progenitors for each host halo ($N_{\rm prog}=\sum_i M^i_{\rm star, dest}/ \langle M_{\rm star, dest} \rangle$, cf.\ \citealt{cooper10}). Generally, $1-2$ dwarfs comprise the majority of the stellar mass contributed by destroyed dwarf galaxies.

\subsection{Dependence on Accretion History}

\begin{figure}
\begin{center}
\includegraphics[width=8.5cm, height=12.75cm]{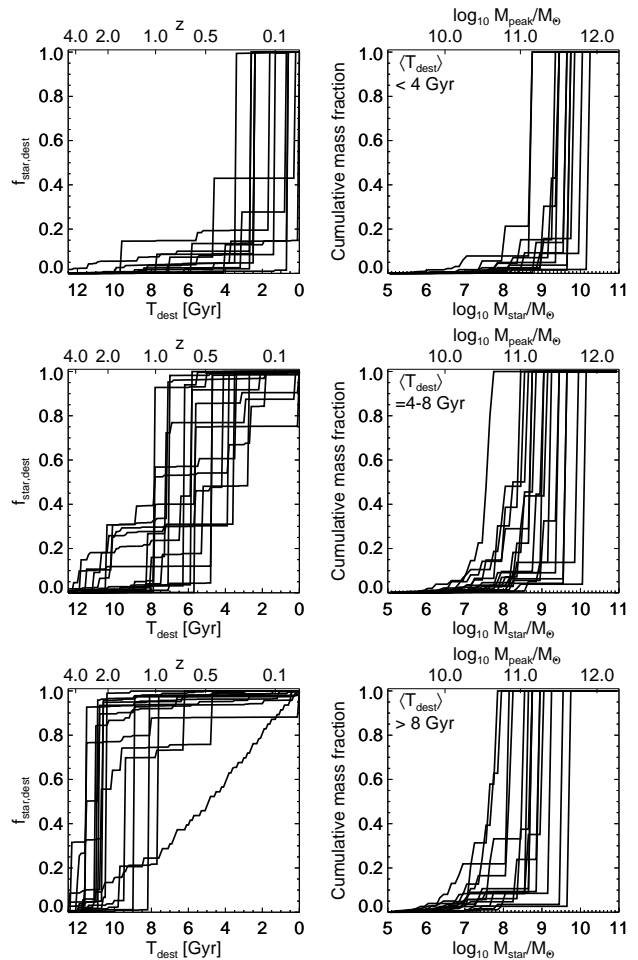}
\caption{\label{fig:mdest_tdest} \small \textit{Left panels:} The cumulative fraction of accreted stellar mass from destroyed dwarfs as a function of lookback time. Each row shows the cumulative fractions for a range $\langle T_{\rm dest} \rangle$, where $\langle T_{\rm dest} \rangle$ is the average time the dwarfs were destroyed, weighted by their stellar mass. Thus, halos with more quiescent accretion histories (or larger $\langle T_{\rm dest} \rangle$) are shown in the bottom row. \textit{Right panels:} The cumulative fraction of accreted stellar mass from destroyed dwarfs as a function of dwarf mass. The rows are split by $\langle T_{\rm dest} \rangle$ in the same way as the left-hand panels. Halos that accrete most of their stellar mass at earlier times, do so with lower mass dwarfs, while the main contributors for halos that accrete most of their stellar mass at later times are generally more massive dwarfs. Note that this distinction between late- and early-forming halos is largely driven by the addition (or absence) of $\sim 1-2$ massive dwarfs at late times.}
\end{center}
\end{figure}

\begin{figure}
\begin{center}
\includegraphics[width=8cm, height=6.4cm]{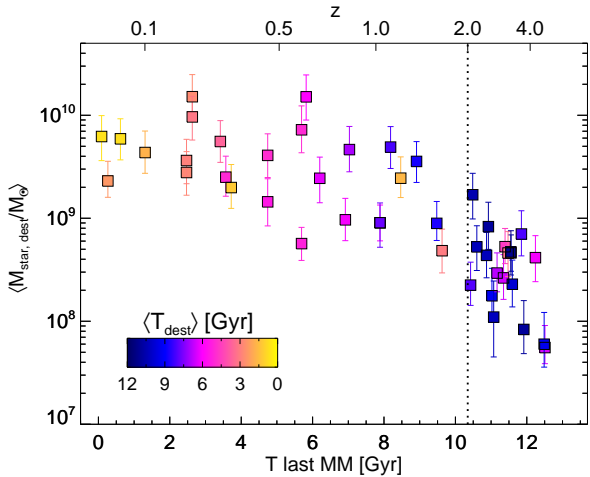}
\caption{\label{fig:last_mm} \small The average, mass weighted, accreted stellar mass from destroyed dwarfs for each halo as a function of the time of its last major merger (dark matter mass ratio $> 0.3$). The error bars show the 1$\sigma$ uncertainty due to the scatter in the $M_{\rm peak}-M_{\rm star}$ relation. The colors indicate the average time the dwarfs were destroyed, weighted by their stellar mass (dark/blue = early times, light/yellow=late times). The dotted line indicates the threshold for our quiescent sample of halos with $z_{\rm LMM} > 2$. Halos with more recent major mergers accrete more massive dwarfs ($M_{\rm star} > 10^9M_\odot$). The accreted stellar material in quiescent halos typically comes from dwarfs with stellar masses of $10^{8}-10^9 M_\odot$.}
\end{center}
\end{figure}

We now consider how the mass spectrum of accreted dwarfs depends on the accretion histories of the host halos. In Figure \ref{fig:mdest_tdest} we show the cumulative fraction of accreted stellar mass from destroyed dwarfs as a function of lookback time (left panels) and dwarf mass (right panels). Each row shows the cumulative fractions for a range $\langle T_{\rm dest} \rangle$.

Halos with earlier accretion epochs and thus earlier $\langle T_{\rm dest} \rangle$ values, tend to build-up their accreted stellar mass from lower mass dwarfs. Halos undergoing recent merger events have larger contributions from more massive dwarfs. Thus, the ``mass-spectrum'' or masses of the most dominant progenitors, depend strongly on the epoch at which most of the stellar mass is accreted. Note that this distinction between late- and early-forming halos is largely due to the addition (or absence) of $\sim 1-2$ massive dwarfs at late times (e.g., recent major mergers).

As shown in Figure \ref{fig:mspec_mstar}, our quiescent sample of host halos are biased towards lower average (mass weighted) destroyed dwarf masses ($\langle M_{\rm star, dest} \rangle$). In Figure \ref{fig:last_mm} we show explicitly how $\langle M_{\rm star, dest} \rangle$ depends on the time of the last major merger of the host halos. Here, we only consider mergers of dwarfs that eventually get destroyed (i.e., we do not include surviving satellites), and we use a (dark matter) mass ratio threshold $> 0.3$ to define major mergers. The filled symbols are colored by $\langle T_{\rm dest} \rangle$. The dotted line indicates the $z=2$ boundary used to define the quiescent sample of halos. As alluded to in the previous section, the quiescent sample have lower $\langle M_{\rm star, dest} \rangle$ values ($\lesssim 10^9 M_\odot$), while halos undergoing more recent major mergers have significant contributions to their accreted stellar mass by more massive dwarfs ($\sim 10^9-10^{10}M_\odot$).

\begin{figure*}
\begin{center}
\includegraphics[width=16cm, height=5.33cm]{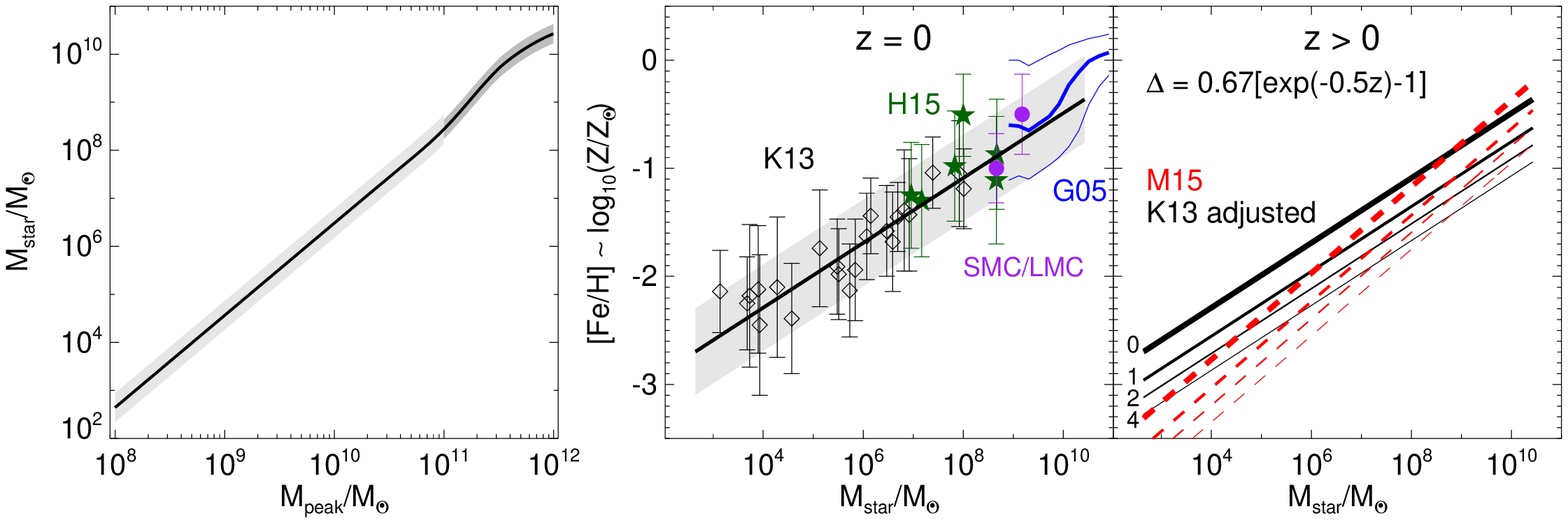}
\caption[]{\label{fig:models} \small \textit{Left panel:} Stellar mass-peak subhalo mass relation adopted in this work. We follow the same prescription as \cite{gk14} assuming 0.2 dex scatter in $\mathrm{log}M_{\rm star}$ at fixed $\mathrm{log}M_{\rm peak}$ for $M_{\rm peak} > 10^{11}M_\odot$, and 0.3 dex scatter for low mass subhalos with $M_{\rm peak} <10^{11}M_\odot$. Our model assumes no redshift evolution in the $M_{\rm star}-M_{\rm peak}$ relation. \textit{Middle panel:} The stellar mass--metallicity relation for Local Group dwarfs (K13: \citealt{kirby13}). The black diamond data points are from \cite{kirby13}. Values for the LMC and SMC are shown with the purple filled circles (\citealt{carrera08}; \citealt{parisi10}), and the green stars show M31 dwarfs from \cite{ho15} (H15). The error bars include the \textit{intrinsic} spread in metallicity for individual dwarfs. The thick blue line shows the relation derived by \cite{gallazzi05} (G05) for local Sloan Digital Sky Survey galaxies. The thinner blue lines show the 16th and 84th percentiles, respectively. A well-defined stellar mass-metallicity relation at $z=0$ exists over 6 orders of magnitude in stellar mass. The approximate standard deviation of the observed metallicity distribution functions for each dwarf is $\sim 0.4$ dex over a wide range in stellar mass (cf.\ \citealt{ho15} Figure 6); this is indicated by the gray band around the mean relation. \textit{Right panel:} The \cite{ma15} (M15) stellar mass--metallicity relation as a function of redshift is shown with the red dashed lines (thicker to thinner lines show $z=0,1,2,4$). We adopt the redshift dependence found in this work ($\Delta [\mathrm{Fe/H}] = 0.67 \left [\mathrm{exp}(-0.5z)- 1\right]$), but we use the $z=0$ \cite{kirby13} relation as the zero-point. The adjusted stellar mass--metallicity--redshift relation is shown with the solid black lines (thicker lines for decreasing redshift).}
\end{center}
\end{figure*}

\section{Metallicity of Destroyed Dwarfs}
\label{sec:met}

In this Section, we consider the metallicity distribution of accreted stellar material contributed by destroyed dwarfs. 

\subsection{Empirical Model}

We employ empirical models to estimate the metallicity of the destroyed dwarfs in the simulations. For completeness, the left-hand panel of Figure \ref{fig:models} shows the stellar mass-peak halo mass relation that we have adopted to assign stellar mass to subhalos (see Section \ref{sec:am}). To relate this stellar mass to an average metallicity ([Fe/H]) we adopt a stellar mass-metallicity relation. Our starting point is the $z=0$ relation for dwarf galaxies, which is well-defined over several orders of magnitude in stellar mass (e.g., \citealt{kirby13}). This relation is shown in the middle panel of Figure \ref{fig:models}; the thick black line shows the best-fit relation derived by \cite{kirby13}. The data points are mainly from the \cite{kirby13} sample (black diamonds). We also show values for massive dwarfs in M31 (\citealt{ho15}, green stars) and the Magellanic Clouds (LMC/SMC, \citealt{carrera08}; \citealt{parisi10}, purple filled circles). The error bars on the observed sample include the \textit{intrinsic} spread in metallicity for individual dwarfs. We also show the relation for slightly more massive dwarfs derived by \cite{gallazzi05} for local Sloan Digital Sky Survey galaxies with the blue lines. The approximate scatter in [Fe/H] at fixed stellar mass is $\sim 0.4$ dex (shown by the gray shaded region). 

At fixed stellar mass, a galaxy at higher redshift has, on average, a lower metallicity than one at lower redshift (e.g., \citealt{erb06}; \citealt{mannucci10}; \citealt{henry13}; \citealt{zahid13}). This redshift evolution is well-studied observationally for $M_{\rm star} \gtrsim 10^{9-10}M_\odot$, but is poorly understood for dwarf galaxies with $M_{\rm star} \lesssim 10^9 M_\odot$. With no observations to guide us, we adopt the relations recently derived by \cite{ma15} from hydrodynamical simulations. These simulations have been successful in matching many observational properties on dwarf mass scales, such as the stellar mass--halo mass relation (\citealt{hopkins14}), the stellar mass-metallicity relation (\citealt{ma15}) and the presence of dark matter cores (\citealt{onorbe15}). The \cite{ma15} redshift evolution of the [Fe/H]--$M_{\rm star}$ relation is shown in the right-hand panel of Figure \ref{fig:models} with the red dashed lines. We adopt the \cite{kirby13} relation at $z=0$, which is slightly shallower than the relation given by \cite{ma15}. However, we use the \textit{redshift evolution} that \cite{ma15} derive: $\Delta$ [Fe/H] $=0.67\left[\mathrm{exp}(-0.5z)-1\right]$. Our adopted stellar mass--metallicity relation for a range of redshifts is shown by the black lines in Figure \ref{fig:models}. Note that our adopted redshift dependence is in good agreement with observational studies probing the metal abundances of $M_{\rm star} > 10^9 M_\odot$ galaxies at different redshifts (e.g. \citealt{rodrigues08}; \citealt{zahid11}).

We note that our assumed $M_{\rm star}-M_{\rm peak}$ relation does not evolve with redshift. Thus, a dwarf surviving today with the same peak subhalo mass as a dwarf that was destroyed $\sim 10$ Gyr ago, is assigned the same stellar mass. This may seem counterintuitive, as the dwarf surviving today could, presumably, have formed more stars. In practice, our simple prescription assumes that both of these dwarfs form the same amount of stars, but over very different timescales, i.e., the dwarf that was destroyed $\sim 10$ Gyr ago formed the same amount of stars as the surviving dwarf, but over a shorter timescale. This simplification, although crude, does naturally take into account the fact that star formation rates are higher at higher redshifts. Furthermore, as noted in Section \ref{sec:am}, we do ensure that our results are not significantly affected if we instead adopt the redshift-dependent $M_{\rm star}-M_{\rm halo}$ relations derived by \cite{behroozi13c} and \cite{moster13}.

To apply our prescription to the subhalos in simulations we must define an appropriate redshift in the subhalo's evolution when we define the metallicity of the dwarf. For example, we could use the redshift when the subhalo reaches its peak mass ($z_{\rm peak}$) or when it first infalls onto the host halo ($z_{\rm infall}$), and assume no star formation occurs after this point. For simplicity, we assign an average metallicity to each dwarf at the redshift it gets destroyed ($z_{\rm dest}$). This is the minimum redshift we could apply, and allows us to naturally agree with the $z=0$ relation for the surviving dwarfs. In practice, it is worth noting that adopting $z_{\rm infall}$ or $z_{\rm peak}$ would lower our derived average metallicities by $\sim 0.2$ dex, but does not significantly affect our conclusions. This is probably because most dwarfs are destroyed relatively rapidly after falling into the host MW halos. 

The appropriate redshift at which star formation ceases likely requires different definitions depending on the stellar mass of the dwarf galaxy. For example, \cite{wetzel15b} and \cite{fillingham15} recently showed that environmental quenching is much more efficient for dwarfs with $M_{\rm star} \lesssim 10^8M_\odot$ than for more massive dwarfs. We find that varying the redshift at which we apply the mass-metallicity relation (e.g., $z_{\rm infall}$ for $M_{\rm star} < 10^8M_\odot$ and $z_{\rm dest}$ for $M_{\rm star} > 10^8M_\odot$) makes little difference to our results. Finally, we note that the $z=0$ ultra-faint dwarf galaxy population likely stopped forming stars several Gyr ago (e.g., \citealt{brown12}), thus the ultra-faint dwarfs destroyed in the past may look very similar (at least in terms of metallicity) to the $z=0$ population. Thus, it may be more appropriate to simply use the observed $z=0$ stellar mass-metallicity relation for these very low-mass dwarfs. However, we find that assuming no redshift evolution in the stellar mass-metallicity relation for dwarfs with $M_{\rm star} < 10^5M_\odot$ does not significantly affect our results. This is because most of the accreted stellar material, even at relatively low metallicity, comes from more massive dwarfs (see Section \ref{sec:fracs}).

In addition to an average metallicity, we assume a Gaussian metallicity distribution function (MDF) for each dwarf with standard deviation of $0.4$ dex, motivated by the observed intrinsic scatter for individual dwarfs at $z=0$. Introducing this intrinsic dispersion is an important component of our model as a more massive dwarf with higher \textit{average} metallicity than a lower mass dwarf can still contribute more stellar mass at lower metallicities owing to the low metallicity tail of its distribution. Of course, the MDFs for individual dwarfs (even at fixed stellar mass) can vary widely (e.g., \citealt{kirby13}; \citealt{ho15}), but our our simple assumption of a Gaussian distribution with fixed dispersion is a good approximation for the ``average'' MDF at a given stellar mass (see below).

In the top panel of Figure \ref{fig:mdf_sig} we show the form of our fiducial Gaussian MDF ($\sigma=0.4$ dex) with the gray filled region. The black dotted line shows a narrower Gaussian with $\sigma=0.2$ dex. The thick purple and green lines show the combined MDFs for luminous dwarf spheroidal galaxies (dSphs) and luminous dwarf irregular galaxies (dIrrs) from \cite{kirby13} (see their Fig. 12). The combined MDFs for the observed dwarfs only include galaxies with $10^6 < M_{\rm star}/M_\odot < 10^8$, and each individual galaxy's MDF was centered at its mean [Fe/H] before the MDFs were stacked together. The average MDF for dSphs is narrower and more peaked than the average MDF for dIrrs. Our model MDF is in good agreement with the average dIrr MDF. The model is less peaked than the average dSph MDF, but is a good approximation to the metal-poor tail of the observed distribution. In contrast, a narrower Gaussian with $\sigma=0.2$ does not agree well with the observed distributions.

\begin{figure}
\begin{center}
\includegraphics[width=8.5cm, height=8.5cm]{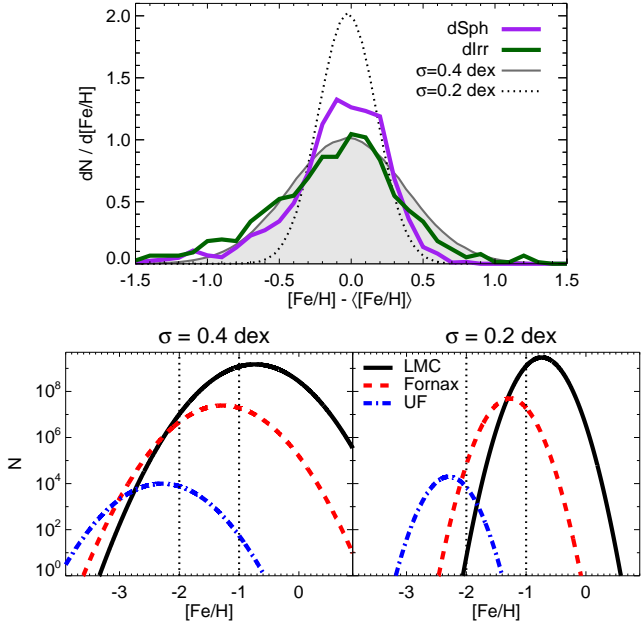}
\caption[]{\label{fig:mdf_sig} \small \textit{Top panel:} We show our Gaussian model metallicity distribution function (MDF) with the gray filled region ($\sigma=0.4$ dex). A narrower Gaussian MDF with $\sigma=0.2$ dex is shown with the dotted black line. Average observed MDFs for dSph and dIrr galaxies ($10^6 < M_{\rm star}/M_\odot < 10^8$) in the local group are shown with the thick purple and green lines, respectively (\citealt{kirby13}). Our fiducial model (with $\sigma=0.4$ dex) agrees well with the average dIrr MDF, and although less peaked, our model is a good approximation to the metal-poor tail of the dSph distribution. \textit{Bottom panels:} Schematic MDFs to show the affect of varying the dispersion in [Fe/H] at fixed stellar mass. We show three Gaussian MDFs of LMC mass ($M_{\rm star} = 1.5 \times 10^9M_\odot$, solid black lines), Fornax mass ($M_{\rm star} = 2.5 \times 10^7M_\odot$, dashed red lines) and ultra-faint mass ($M_{\rm star}=1 \times 10^4 M_\odot$, blue dot-dashed lines) dwarfs, respectively. The average metallicities are chosen from the $z=0$ stellar mass-metallicity relation (\citealt{kirby13}). The y-axis is logarithmic in order to compare masses over 5 orders of magnitude. The contributions from more massive dwarfs to the metal-poor component is strongly related to the metal-poor tails of their MDFs. If the MDFs are narrow (right-panel) dwarfs with high average metallicities will contribute very little to the metal-poor component.}
\end{center}
\end{figure}

\begin{figure*}
\begin{center}
\includegraphics[width=16cm, height=5.33cm]{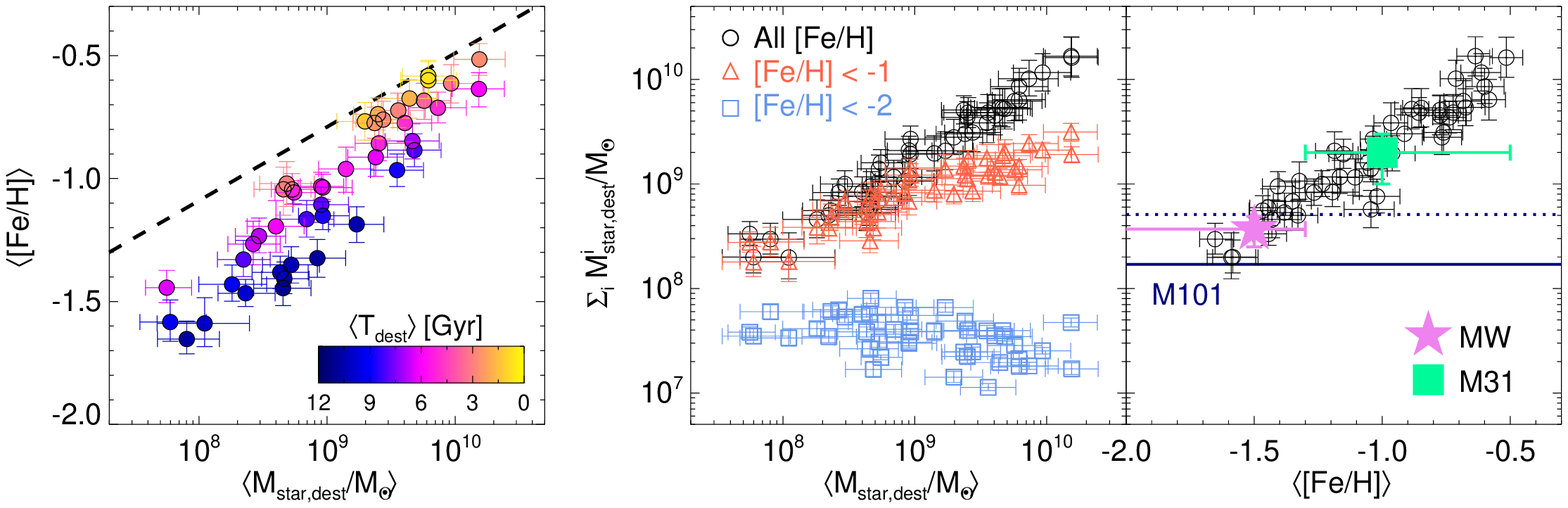}
\caption[]{\label{fig:met_tot} \small \textit{Left panel:} The average metallicity of accreted stellar material in each halo as a function of the mass-weighted average destroyed dwarf mass. The points are color coded according to the mass weighted average time of destruction. This relation reflects the stellar mass-metallicity-redshift model that we have adopted. The dashed black line shows the $z=0$ stellar mass-metallicity relation from \cite{kirby13}. \textit{Middle panel:} The total accreted stellar mass as a function of the mass weighted average destroyed dwarf mass. Halos built up from more massive dwarfs have more massive accreted stellar components at $z=0$. The red triangles and blue squares show the total accreted metal-poor ([Fe/H] $<-1$) and very metal-poor ([Fe/H] $<-2$) stellar mass, respectively. The total mass of the very metal-poor component of the accreted stars is approximately independent of the typical mass dwarf that built-up the halo. \textit{Right panel:} The total accreted stellar mass as a function of the average metallicity of the accreted material from destroyed dwarfs. The pink star and green square symbols indicate approximate observed values for the MW and M31, respectively. The solid navy line shows the estimated stellar halo mass for the for the nearby galaxy M101, and the dotted navy line shows the 1$\sigma$ upper limit.}
\end{center}
\end{figure*}

In the bottom panels of Figure \ref{fig:mdf_sig} we show schematically the affect of varying the dispersion of the MDFs. Here, we adopt the $z=0$ mass-metallicity relation (\citealt{kirby13}), and show Gaussian MDFs for LMC mass ($M_{\rm star} = 1.5 \times 10^9M_\odot$, solid black lines), Fornax mass ($M_{\rm star} = 2.5 \times 10^7M_\odot$, dashed red lines) and ultra-faint mass ($M_{\rm star}=1 \times 10^4 M_\odot$, blue dot-dashed lines) dwarfs. The dispersion in [Fe/H] we adopt directly affects the contribution of different mass dwarfs to the metal-poor component of the accreted stars (see Section \ref{sec:fracs}). Most importantly, \textit{the metal-poor tails} of the MDFs for more massive dwarfs are vital. Below, we outline our results with our fiducial assumption of Gaussian metallicity distributions with $\sigma=0.4$ dex, but we also comment on how our results are affected by our adopted $\sigma$.

\subsection{Metallicity dependence of the Mass Spectrum}

After applying our empirical models (described above), we can investigate the approximate metallicity of the stellar material accreted by destroyed dwarfs.

The left-hand panel of Figure \ref{fig:met_tot} shows the average (mass weighted) metallicity of the accreted stellar component in each host halo against the average (mass weighted) destroyed dwarf stellar mass ($\langle M_{\rm star, dest} \rangle$). The colors indicate the average (mass weighted) lookback time when these dwarfs were destroyed. This relation reflects our adopted stellar mass-metallicity-redshift relation, whereby more massive dwarfs have higher metallicities, and dwarfs destroyed at earlier times have lower metallicities.

We can use the average metallicities of the destroyed dwarfs, and their intrinsic scatter (0.4 dex) to estimate the fraction of the total accreted stellar mass that is low metallicity ([Fe/H] $<-1$) or very low metallicity ([Fe/H] $< -2$). This is shown in the middle panel of Figure \ref{fig:met_tot} as a function of $\langle M_{\rm star, dest} \rangle$. The total stellar mass (black circles) and total low metallicity ([Fe/H] $<-1$, red triangles) stellar mass increases with $\langle M_{\rm star, dest} \rangle$. However, the total stellar mass of very low metallicity stars ([Fe/H] $<-2$, blue squares) stays approximately constant with $\langle M_{\rm star, dest} \rangle$. Thus, regardless of the average (mass weighted) mass of the destroyed dwarfs, approximately the same mass of very low metallicity stars is accreted by each halo.

In the right-hand panel of Figure \ref{fig:met_tot} we show the total accreted stellar mass against the average metallicity of the accreted material. For comparison, we show the approximate observational constraints for the stellar halos of the MW ($M_{\rm star, halo} \sim 3.7 \pm 1.2 \times 10^8M_\odot$; \citealt{bell08}, [Fe/H] $\sim -1.3$ to $-2.2$; \citealt{carollo10}), M31 ($M_{\rm star, halo} \sim 2 \pm 1 \times 10^9M_\odot$; \citealt{williams15},  [Fe/H] $\sim -0.5$ to $-1.3$; \citealt{kalirai06}), and the nearby galaxy M101 ($M_{\rm star, halo} =  1.7^{+3.4}_{-1.7} \times 10^8M_\odot$; \citealt{vandokkum14}). The agreement is pretty remarkable, especially given the simplicity of our models. However, it is worth cautioning that our accreted stellar masses are only an approximate representation of the \textit{stellar halos} of galaxies. This is because some of the accreted material can end up in the disk/bulge and a significant fraction of the observed masses could be contributed by halo stars born in-situ (e.g., \citealt{zolotov09}; \citealt{font11}). Nonetheless, it is reassuring that our simple empirical models provide a reasonable agreement with the available observational constraints, and, perhaps more importantly, we can reproduce the \textit{relative} difference between the MW and M31 stellar halos. This difference is, at least in part, likely indicative of their respective accretion histories. We also note that the models by \cite{purcell08} find diffuse intrahalo light metallicities of $\mathrm{log}_{10} Z/Z_\odot \sim -1.0 \pm 0.5$ for $10^{12}M_\odot$ mass host halos, in good agreement with our models.

The stellar halo masses and average metallicities of the MW and M31 halos can be used to roughly estimate the dominant contributors to their accreted stellar components. The MW seems to favor relatively low-mass progenitors $\langle M_{\rm star, dest} \rangle \sim 10^8M_\odot$ (SMC/Sagittarius mass), while M31 is more consistent with an order of magnitude larger progenitors $\langle M_{\rm star, dest}\rangle \sim 10^9M_\odot$ (LMC mass). Note that the very low stellar halo mass recently measured by \cite{vandokkum14} for M101 suggests relatively low-mass progenitors (similar to the MW or lower) and also predicts a low average metallicity. While we only make rough comparisons with observations here, our simple models show that the dominant halo progenitors can be inferred from measurements of the stellar mass in the halo of galaxies and/or their average metal content. Both of these observational measurements are, or will be, feasible for many MW-mass galaxies in the nearby Universe.

\subsection{Contributions from ``Classical'' and ``Ultra-Faint'' Dwarfs}
\label{sec:fracs}
\begin{figure}
\begin{center}
\includegraphics[width=8cm, height=12cm]{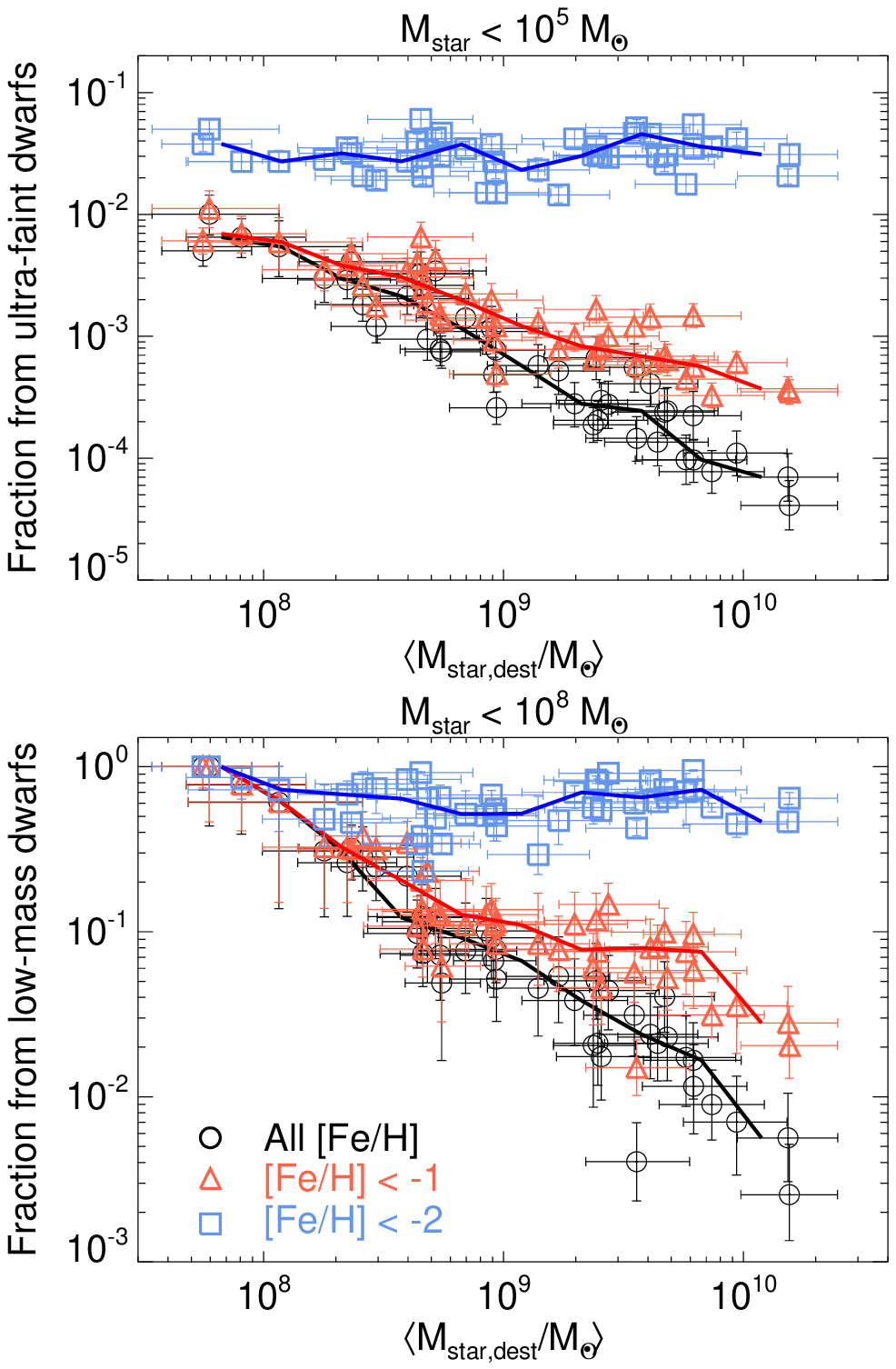}
\caption{\label{fig:met_frac} \small The fractional contribution to the total accreted stellar mass from ultra-faint dwarfs ($M_{\rm star} < 10^5 M_\odot$, top panel) and low-mass dwarfs ($M_{\rm star} < 10^8 M_\odot$, bottom panel) as a function of the average mass weighted stellar mass of destroyed dwarfs, $\langle M_{\rm star, dest} \rangle$. The black circles, red triangles and blue squares show the fractions for the accreted stellar component with all [Fe/H], metal-poor [Fe/H] $<-1$ and very metal-poor [Fe/H] $<-2$, respectively. The solid lines indicate the median values over all 45 host halos in (0.25 dex) bins of $\langle M_{\rm star, dest} \rangle$. Ultra-faint dwarfs contribute very little accreted mass at all metallicities. Even for the most metal-poor component ([Fe/H] $<-2$) the contribution from ultra-faints is small ($\sim 2-5\%$). Most of the stellar mass comes from dwarfs with $M_{\rm star} > 10^8M_\odot$, however low-mass dwarfs ($10^5 < M_{\rm star}/M_\odot < 10^8$) contribute a significant amount ($\sim 40-80\%$) to the very low metallicity component.}
\end{center}
\end{figure}

We now use the empirical models developed in the previous section(s) to estimate the contributions from low mass dwarfs ($M_{\rm star} < 10^8M_\odot$) and ultra-faint dwarfs ($M_{\rm star} < 10^5M_\odot$) to the accreted stellar material.

We also consider the contributions from different mass dwarfs to the metal-poor accreted stellar material. Naively one may expect that the most metal-poor stars come from the the lowest mass dwarfs with the lowest \textit{average} metallicity. However, although the average metallicities of more massive dwarfs are higher than low mass dwarfs, the \textit{tail} of their MDFs can still contribute more metal-poor stars because they contain many more stars (see Figure \ref{fig:mdf_sig}). 

Figure \ref{fig:met_frac} shows the fractional contribution from ultra-faint (top panel) and low-mass dwarfs (bottom panel) as a function of the average (mass weighted) destroyed dwarf stellar mass ($\langle M_{\rm star, dest} \rangle$). The black circles are for the overall stellar mass and the red triangles and blue squares show the metal-poor ([Fe/H] $< -1$) and very metal-poor ([Fe/H] $<-2$) components, respectively. 

Ultra-faint dwarfs contribute very little to the accreted stellar mass regardless of the mass spectrum of destroyed dwarfs. Even at very low metallicities, they only contribute $\sim 2-5\%$ of the accreted stellar mass. On the other hand, low-mass ``classical'' dwarfs ($10^5 < M_{\rm star}/M_\odot < 10^8$) can contribute a significant amount to the overall accreted material if $\langle M_{\rm star, dest} \rangle$ is low. Furthermore, regardless of the mass-spectrum, they contribute a significant amount of the very metal-poor stars ($\sim 40-80\%$). It is worth remarking that more massive dwarfs ($M_{\rm star} > 10^8M_\odot$) can still contribute a significant amount ($\sim 20-60 \%$) to the very metal-poor material, even though their average metallicities are [Fe/H] $\gg -2$. Furthermore, they are generally the main contributors to the stellar material with $-2<$ [Fe/H] $<-1$.

As shown in Figure \ref{fig:mdf_sig}, the intrinsic scatter of the dwarf MDFs is important when considering the contributions to the metal-poor component of the accreted stars. We find that adopting a narrower dispersion ($\sigma =0.2$ dex) has little affect on the contributions to the accreted material with [Fe/H] $> -2$, but has a non-negligible affect at the lowest metallicities. With less intrinsic scatter, more massive dwarfs ($M_{\rm star} > 10^8M_\odot$) contribute significantly less stellar material with [Fe/H] $<-2$, and the majority ($\gtrsim 95 \%$) of the lowest metallicity stellar mass comes from dwarfs with $M_{\rm star} < 10^8M_\odot$. However, the contribution from ultra-faint mass dwarfs is still not dominant, with typical fractions of $\sim 10\%$ and a maximum fraction of $\sim 20\%$. 

It is worth re-emphasizing that our adopted model assumes that \textit{all} subhalos with $M_{\rm peak} > 10^8M_\odot$ host luminous galaxies. Given that it is likely that a significant fraction of subhalos with $M_{\rm peak} \lesssim 10^{9.5}M_\odot$ do not form any stars (see Section \ref{sec:am}), we are likely \textit{overestimating} the contribution to the accreted stellar material from the ultra-faint mass dwarfs.

Finally, we note that we cannot rule out that ultra-faint dwarfs may be major contributors to galaxy stellar halos at large radii, but their contribution in the inner regions of the halo, where the majority of the stellar halo mass resides, is likely piddly. Furthermore, they may be the dominant source of rare, extremely metal-poor stars in galaxy halos with [Fe/H] $\ll -3$.

\section{Surviving and Destroyed Dwarfs}
\label{sec:sats}

In this final section, we consider the relation between the surviving dwarf population at $z=0$ and the dwarfs that have since been destroyed.

\begin{figure}
\begin{center}
\includegraphics[width=8cm, height=12cm]{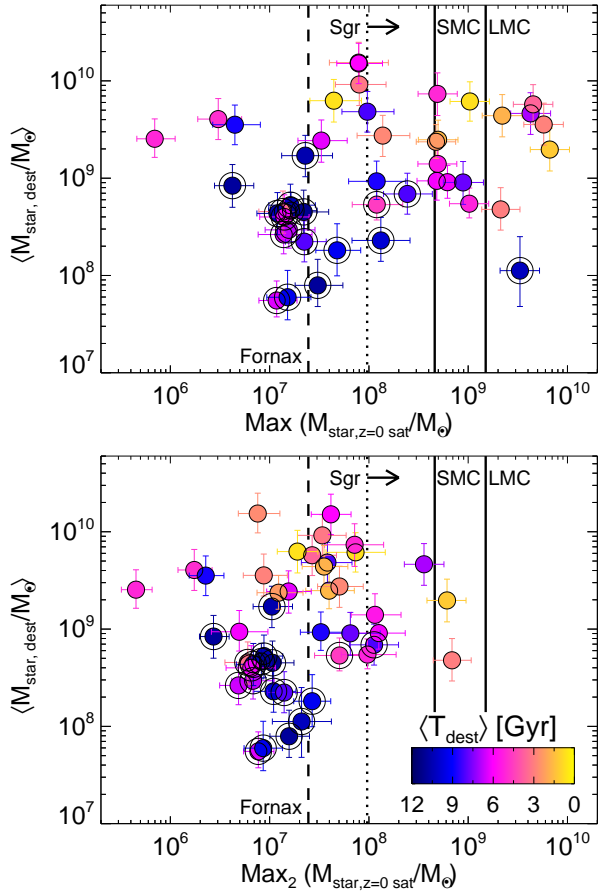}
\caption{\label{fig:msat} \small The typical mass dwarf that contributes to the accreted stellar mass as a function of the most massive (top panel) and second most massive (bottom panel) surviving dwarfs at $z=0$. The filled symbols are color coded according to the mass weighted average time of destruction for destroyed dwarfs (blue/dark = early times, yellow/light= late times). The approximate stellar masses of the LMC, SMC, Sagittarius (Sgr) and Fornax are shown with the vertical lines. Note that the stellar mass for Sgr is a lower limit (\citealt{NO10}). Halos that have accreted relatively massive dwarfs ($M_{\rm star} > 10^9M_\odot$) have a range of surviving satellite masses. However, halos that have only accreted lower mass dwarfs ($M_{\rm star} < 10^9M_\odot$) tend to have less massive surviving satellites today. The encircled symbols indicate the quiescent halo sample. Only one of the quiescent halo sample has a satellite as massive as the LMC, but the second most massive satellite of this halo is significantly less massive than the SMC or Sgr dwarf. This indicates some tension between the generic assumption of a ``quiescent'' MW halo and the mass spectrum of its surviving satellites.}
\end{center}
\end{figure}

Figure \ref{fig:msat} shows the average (mass weighted) stellar mass of destroyed dwarfs against the mass of the most massive (top panel) and second most massive (bottom panel) surviving dwarf at $z=0$. The filled symbols are color coded according to the mass weighted average time of destruction for destroyed dwarfs. The black circles indicate the subset of halos with quiescent accretion histories. Solid lines indicate the approximate stellar masses of the Large and Small Magellanic clouds (LMC/SMC, \citealt{mcconnachie12}), and the dotted line shows the approximate lower limit on the Sagittarius dwarf galaxy stellar mass derived by \cite{NO10}. We also indicate the stellar mass of the Fornax dwarf, the most massive classical dwarf, with the dashed black line.

Halos that typically destroyed less massive dwarfs ($\lesssim 10^9 M_\odot$), tend to have less massive surviving satellites today. Halos that have destroyed more massive dwarfs tend to have more massive surviving satellites, but there is a lot of scatter.

Only one ($\sim 6\%$) of the quiescent halo sample has a surviving dwarf with a mass comparable to the LMC ($M_{\rm star} > 10^9M_\odot$), compared to 30\% for the non-quiescent sample. Moreover, \textit{none} of the quiescent halos have an LMC-mass dwarf at $z=0$ and a second most massive dwarf with $M_{\rm star} > 10^8 M_\odot$. This suggests some potential tension between the general assumption of a quiescent MW mass halo and the mass spectrum of its surviving $z=0$ dwarfs.
Note that 20\% of the overall sample of MW-mass halos at $z=0$ host a dwarf satellite with $M_{\rm star} > 10^9M_\odot$, and 7\% host a second most massive satellite with $M_{\rm star} > 2 \times 10^8M_\odot$. These fractions are in good agreement with the numbers of LMC and SMC analogues in $10^{12}M_\odot$ halos found by \cite{boylan11} and \cite{busha11} using the Millennium II and Bolshoi simulation suites, respectively. Observational studies by \cite{liu11}, \cite{tollerud11} and \cite{robotham12} using spectroscopic samples from the Sloan Digital Sky Survey and the Galaxy And Mass Assembly project, also find comparable fractions of LMC- and SMC-mass satellites to these simulations\footnote{These studies use different selection criteria, though identical criteria were used in comparing the simulated study of \cite{busha11} to the observational study of \cite{liu11}}. Thus, although our results are based on a relatively small number of host halos, our statistics of massive satellites are in agreement with larger samples in both simulations and observations.

The mass spectrum of surviving dwarfs today can be compared to the ``mass-gap'' statistic often used on group/cluster scales to classify fossil groups (e.g., \citealt{ponman94}; \citealt{jones03}). Here, halos with more massive satellites, and thus smaller logarithmic ``gaps'' between the host halo mass and most massive satellite mass, tend to have younger and less concentrated dark matter halos. Fossil groups have large mass-gaps and tend to be old and highly concentrated (e.g., \citealt{donghia05}; \citealt{beckmann08}; \citealt{dariush10}; \citealt{deason13b}). Again, scaling these relations down to MW mass scales presents somewhat of a conundrum, as the MW halo is likely old and perhaps even highly concentrated (e.g., \citealt{battaglia05}; \citealt{smith07}; \citealt{deason12}), despite having a massive LMC satellite. However, this extrapolation completely ignores the scatter in the mass-gap---halo age relation, which can be considerable. For example, \cite{deason13b} showed that a significant fraction ($\sim$ 20\%) of groups with large mass-gaps are young, and likely experienced a recent major merger between a massive satellite subhalo and the central subhalo (cf.\ the halo in Figure \ref{fig:msat} with $\langle M_{\rm star, dest} \rangle \sim 10^{10}M_\odot$, $\langle T_{\rm dest} \rangle \sim 4$ Gyr, and Max($M_{\rm star, z=0 \, sat}$) $\sim 10^6M_\odot$). Conversely, there are halos that have recently accreted a massive satellite subhalo, but have had little ``action'' prior to this event. Thus, the transient nature of the halo mass-gap statistic leads to a population of halos that can be labeled as ``transient fossils''; these are halos with a recent merger or accretion event that masks the preceding formation history of the halo. Given the observational evidence that the LMC and SMC were probably accreted very recently (e.g., \citealt{besla07}; \citealt{kalli13}), it seems likely that the MW is one of these so-called transient fossils.

The ``uniqueness'' of the MW is an important topic to address, as our Galaxy is often viewed as a benchmark $L^\star$ galaxy that we use to compare with both observations of external galaxies and simulations. It is well-known that less massive MW halos are less likely to host LMC/SMC mass satellites at $z=0$ (e.g., \citealt{boylan10}). Thus, the combination of a relatively low MW halo mass ($\sim 10^{12} M_\odot$) and a quiescent accretion history would suggest that the MW is even more of an oddity than previously thought. The presence of an LMC/SMC has been used to probabilistically determine the mass of the MW using large samples of halos in numerical simulations (e.g., \citealt{busha11}; \citealt{catun14}). Our results suggest that the inclusion of a proxy for accretion history in such calculations could have a significant affect on these inferences.

\section{Conclusions}
\label{sec:conc}
We used a suite of 45 zoom-in simulations of MW mass halos to study the mass spectrum of destroyed dwarfs that contribute to the accreted stellar mass of the halos. The halos have a narrow mass range, $M_{\rm vir}=10^{12.1\pm 0.03}M_\odot$, which allows us to focus on the variation in assembly histories of the host halos. Empirical models are used to relate (peak) subhalo mass to stellar mass, and we use constraints from hydrodynamical simulations as well as $z=0$ observations to estimate the metallicity distribution of the accreted stellar material. Our main conclusions are summarized as follows:

\begin{itemize}

\item Typically, $1-2$ destroyed dwarfs with stellar masses of $10^8-10^{10}M_\odot$ contribute the majority of the accreted stellar mass of MW mass halos. The mass-weighted average stellar masses of destroyed dwarfs are strongly related to the assembly history of the host halos. The accreted stellar mass of host halos with quiescent histories are built up from lower mass dwarfs ($10^8-10^9M_\odot$) and have lower total accreted stellar masses at $z=0$. Halos undergoing recent major mergers have larger total accreted stellar masses, and are dominated by more massive destroyed dwarfs ($\gtrsim 10^9M_\odot$). The dominant contributors to the accreted stellar mass are, for all halos, relatively high mass ($\gtrsim 10^8M_\odot$) dwarfs due to the steep relation between $M_{\rm star}$ and $M_{\rm peak}$ at low halo masses. 

\item The average metallicity of the accreted stellar material reflects the mass spectrum of the halo progenitors, as well as the time at which these dwarfs were destroyed. Low mass dwarfs destroyed at early times are lower metallicity than higher mass dwarfs destroyed relatively recently. Our derived relation between the average metallicity of accreted stellar mass and the total accreted stellar mass at $z=0$ is in good agreement with observational constraints for the stellar halos of the MW and M31. Accreted components with lower average metallicity and lower total mass, are likely present in host halos with more quiescent accretion histories. Thus, the higher average metallicity and total stellar halo mass of M31 relative to the MW suggests a more active, recent accretion history for the M31 galaxy. We note that employing stringent constraints on the assembly histories to select ``MW-type'' galaxies (i.e., no major mergers since $z=2$) can introduce severe biases, and likely excludes several disk galaxies (like M31) with similar masses to the MW.

\item The contribution to the total accreted stellar mass from classical mass ($10^5 < M_{\rm star}/M_\odot < 10^8$) and ultra-faint mass ($M_{\rm star} < 10^5M_\odot$) dwarfs depends on the destroyed dwarf mass spectrum, and thus the host halo assembly history.  For all halos, the contribution from ultra-faint dwarfs is negligible ($\ll 1 \%$). However, classical dwarfs can contribute a significant amount to the accreted stellar mass in halos with very quiescent accretion histories. Furthermore, regardless of the mass spectrum, classical dwarfs contribute a substantial amount ($\sim 40-80 \%$) to the accreted very metal-poor stars ([Fe/H] $< -2$). On the other hand, although the low mass ultra-faint dwarfs have lower average metallicities, their contribution to the very metal-poor stellar material is low ($\sim 2-5\%$). In fact, if more massive dwarfs ($M_{\rm star} > 10^8M_\odot$) have significant metal-poor tails to their metallicity distributions, they can contribute a considerable amount ($\sim 20-60\%$) to the very metal-poor material even though their average metallicities are [Fe/H] $\gg -2$. Furthermore, these more massive dwarfs are generally the main contributors to the stellar material with $-2 <$ [Fe/H] $< -1$. 

\item By comparing the average (mass weighted) destroyed dwarf mass to the surviving $z=0$ satellite population, we find that halos with relatively low mass progenitors, and thus relatively quiescent accretion histories, tend to have lower mass surviving dwarfs today. We find that only one of the ``quiescent'' host halos has a surviving satellite with mass similar to the LMC, and \textit{none} also have a second most massive satellite of similar mass to the SMC and/or Sagittarius. Thus, the generic assumption of a quiescent MW halo seems in tension with the mass spectrum of its surviving dwarfs. We suggest that the MW could be a ``transient fossil'' --- a quiescent halo with a recent accretion event(s) that disguises the preceding formation history of the halo.

\end{itemize}

Our analysis combines high-resolution $\Lambda$CDM dark matter only simulations with empirical galaxy formation models (i.e., the stellar mass--halo mass relation and the stellar mass--metallicity--redshift relation) to study the stellar material accreted by MW-mass halos from destroyed dwarfs. Our results are a natural outcome of some of these underlying assumptions. For example, the relation between stellar mass and halo mass is steep for dwarf galaxies ($M_{\rm star} \propto M_{\rm peak}^{1.9}$), and despite their abundance, this steep relation diminishes the contribution of very low-mass dwarfs to the overall accreted stellar mass. It is worth emphasizing that the ``true'' relation between stellar mass and halo mass for low-mass galaxies is highly uncertain, but, as far as we are aware, there is no observational or theoretical evidence for a significantly flatter $M_{\rm star}-M_{\rm halo}$ relation that would alter our conclusions. 

Under our model assumptions, the dominant contributors to the accreted stellar mass are LMC/SMC-mass dwarfs ($\sim 10^9 M_\odot$), and ``classical'' mass dwarfs ($10^5 < M_{\rm star}/M_\odot < 10^8$) generally supply the majority of the very metal-poor stellar material. In Section \ref{sec:met}, we highlighted the importance of our adopted MDFs of dwarf galaxies on our results, particularly for the more massive dwarfs. Although our models are chosen to reflect the current observational (and theoretical) MDFs of dwarfs, observations are somewhat scarce, and ultimately tied only to local group dwarfs. This work could greatly benefit from larger statistical samples of dwarf galaxy MDFs, over a range of masses and redshifts. At present, this is a daunting task, however the prospect of upcoming observational facilities, such as the 30-m class telescopes and wide-field spectrographs on 10-m class telescopes, will greatly facilitate this goal in the near future.

\acknowledgments{
AJD is supported by a Porat Fellowship at Stanford University.
We thank Evan Kirby and Andrew Wetzel for stimulating science discussions, and for providing very useful comments on an earlier version of this manuscript. We also thank an anonymous referee for providing valuable comments that improved the quality of our paper.
The simulations used were run using computational resources at SLAC; we gratefully acknowledge the support of the SLAC computational team.
This research also used resources of the National Energy Research Scientific Computing Center, a DOE Office of Science User Facility supported by the Office of Science of the U.S. Department of Energy under Contract No.~DE-AC02-05CH11231.
AJD thanks JBC for his impeccable timing and invaluable insight.
}

\bibliographystyle{yahapj}
\bibliography{mybib}

\end{document}